# Optically Induced Magnetization of CdMnTe Self-Assembled Quantum Dots

S. Mackowski*, T. Gurung, T.A. Nguyen, H.E. Jackson, and L.M. Smith

*Department of Physics, University of Cincinnati, 45221-0011 Cincinnati OH*

G. Karczewski and J. Kossut

*Institute of Physics Polish Academy of Sciences, 02-668 Warszawa, Poland*

We demonstrate that resonant excitation of CdMnTe self-assembled quantum dots creates an ensemble of spin-polarized magnetic polarons at B=0 T. The strong spatial confinement characteristic of quantum dots significantly increases the stability of magnetic polarons so that the optically induced spin alignment is observed for temperatures > 120 K.

*Author to whom the correspondence should be addressed: electronic mail: seb@physics.uc.edu

A central part of the effort to take advantage of carrier spin for new electronic devices is to develop techniques to control and manipulate the spin alignment of magnetic impurities in diluted magnetic semiconductor (DMS) nanostructures [1]. Recently, carrier-induced ferromagnetism has been demonstrated for GaMnAs epilayers [2] and p-type CdMnTe quantum wells (QWs) [3]. Curie temperatures of 120K and 3K, respectively, have been achieved. The holes which stabilize ferromagnetic alignment of Mn ions were provided either by Mn acting as acceptor [2] or through modulation doping with nitrogen [3]. Another approach applied earlier to HgMnTe [4] and undoped CdMnTe QWs [5] used suitably polarized optical excitation to align magnetic impurities. The optically induced magnetization was detected using a SQUID magnetometer [4] or the induced current in a pickup-coil [5]. These experiments have shown that for CdMnTe QW the magnetization is built through formation of exciton magnetic polarons (EMPs) [5], where the magnetic impurities align spontaneously within the Bohr radius of the exciton in order to minimize the energy. Unfortunately, due to instability of the EMPs and the very rapid time decay of this spin alignment (~100ps [6]), this effect was only observed in pulsed optical experiments at low temperatures. A theoretical model proposed by Dietl *et al.* has interpreted this rapid demagnetization as a result of spin-spin interaction within the EMP [7].

Incorporation of magnetic ions into semiconductor QDs offers the possibility of studying the interaction between carriers and magnetic ions under strong spatial electronic confinement [8-10]. It has been shown that EMPs confined in magnetic QDs, show significant enhancement of their thermal stability [11]. In addition, the spin relaxation time in non-magnetic semiconductor QDs has been found to be considerably longer than in QWs, reaching several hundreds of picoseconds [12,13]. One might thus anticipate that it might be possible to polarize Mn spins within a DMS QD by a suitably polarized exciton.

In this letter we demonstrate the observation of optically induced magnetization of the Mn spins embedded in CdMnTe QDs. We show that through the resonant excitation of spin-polarized excitons one can control the alignment of zero-dimensional EMPs in QDs and therefore the magnetization direction of large QD ensembles. Moreover, due to significant enhancement of the stability of spin-polarized EMPs in these zero-dimensional nanostructures this optically induced spin alignment is observed above 120 K.

The samples containing magnetic CdMnTe QDs and non-magnetic CdTe QDs were grown by molecular beam epitaxy on (100)-oriented GaAs substrates. Prior to the growth of the QD layer, thick CdTe and ZnTe buffer layers were grown sequentially at a temperature of $350^0$C. In order to obtain magnetic QDs, Mn ions were incorporated by exposing the ZnTe surface to Mn flux for 2 seconds before depositing the CdTe QDs layer [9]. The QDs themselves were formed after depositing 4 monolayers of CdTe by atomic layer epitaxy. Finally, the dots were capped by a 50-nm thick ZnTe layer. The presence of Mn within the QDs has been demonstrated by single dot spectroscopy and magneto-photoluminescence data [9].

Spin-polarized excitons were created resonantly in the QDs by LO phonon-assisted absorption into the QD ground states by both $\sigma^+$ - and $\sigma^-$- polarized light [14]. Continuous wave argon ion-pumped dye laser (Rhodamine 590) was used as a tunable excitation source. The sample was placed in a variable temperature (from 4 K to 120 K) continuous-flow helium cryostat. Polarization of both excitation and emission was controlled by Babinet-Soleil compensators and Glan-Thomson linear polarizers, which enables the measurement of the emission intensity in both circular polarizations as a function of the excitation polarization. The emission was dispersed by a triple monochromator and detected by a cooled CCD camera.

In Fig. 1a we show resonantly excited photoluminescence (PL) spectra for non-magnetic CdTe QDs measured at B=0 T. If one excites the inhomogeneously broadened QD ensemble resonantly, spectral lines are observed that are significantly sharper than the non-resonant spectrum (shown by the shaded region in Fig. 1a). These lines are related to LO phonon-assisted absorption in QDs [14]. It is important to note that spin-polarized excitons are excited *directly* into the ground states of QDs responsible for this enhanced emission. For the resonant spectra shown in Fig. 1a the excitation is $\sigma^+$ - polarized while we analyze both $\sigma^+$ (circles) and $\sigma^-$ (squares) - polarized emission. We find that in the case of the non-magnetic CdTe QDs the integrated PL intensity and its shape for the $\sigma^+$ and $\sigma^-$ – polarized emissions are *identical*, indicating that initially spin-polarized excitons randomize their spins within a time much less than the exciton recombination time (~300 ps) [15]. We observe the same result for $\sigma^-$ – polarized excitation, suggesting again a rapid scattering of excitons between the degenerate spin states before recombination.

When Mn ions are placed into a II-VI semiconductor, exchange interactions between the electrons and holes and the d-electrons localized on the Mn ions result in a large sensitivity of the exciton spin to the magnetic field [16]. Indeed, the energy splitting between the exciton spin states - being directly proportional to the Mn magnetization - can be much larger than the thermal energy or the exciton binding energy. Moreover, the exchange energy enables the exciton to lower its energy at zero field by forming EMP through spontaneous alignment of the magnetic ions within its Bohr radius [16]. Each single EMP is aligned along an arbitrary direction, which minimizes its energy and is determined by the particular local distribution of magnetic impurities.

Therefore, it is remarkable that when spin-polarized excitons are resonantly photo-excited into the magnetic QDs significant spin polarization of zero-dimensional EMP emission is observed at B=0T.  In Fig. 1b we show resonantly excited PL spectra for the CdMnTe QDs taken under the same conditions as for the CdTe QDs shown in Fig. 1a.  It is clear that the PL response of the magnetic QDs is *significantly* different.  We find that resonant $\sigma^+$ – polarized excitation leads to predominantly $\sigma^+$ – polarized emission. Moreover, for $\sigma^-$ – polarized excitation the QD emission is also strongly $\sigma^-$ – polarized.  These results indicate that spin-polarized excitons create a significant spin polarization of EMPs in DMS QDs at B=0 T.

This behavior is seen even more clearly in the plots in Fig. 2 that show polarization as a function of emission energy for both $\sigma^+$ – and $\sigma^-$ – polarized excitations as the temperature is increased from 4 K to 120 K.  The polarization of QD emission is calculated from spectra similar to those shown in Fig. 1 using the equation $P(\varepsilon) = (I_+(\varepsilon) - I_-(\varepsilon))/(I_+(\varepsilon) + I_-(\varepsilon))$.  The emission of CdTe QDs shows no polarization for any emission energy (solid line in Fig. 2).  In contrast, we observe strong (40% for the first LO phonon resonance) positive polarizations for $\sigma^+$ – polarized excitation and strong negative polarizations for $\sigma^-$ – polarized excitation for the magnetic CdMnTe QDs at T=4 K.  With increasing temperature the optically induced spin alignment of the EMPs in the CdMnTe QDs decreases monotonically but persists for T>120 K.

Two important conclusions may be drawn from these results:  (1) through polarized resonant excitation, we are able to build up a significant spin polarization of the EMPs within the CdMnTe QDs, and (2) due to strong spatial confinement in QDs, the spin polarization of EMPs is robust and persists up to T=120 K.

The mechanism responsible for the large PL polarization at *zero magnetic field* observed for the magnetic CdMnTe QDs is illustrated in Fig. 3.  In the absence of an exciton within the magnetic QD, the Mn ions are paramagnetic (Fig. 3a).  If a spin-polarized exciton is photo-excited (Fig. 3b), the Mn ions align themselves along the same direction through formation of an EMP (Fig. 3c). The polarization of the emission should then reflect the actual configuration of magnetic impurities within the dot (Fig. 3d). Recall from the PL results at 4 K (Fig. 1b) that the polarization of Mn ions in CdMnTe QDs is determined by the spin polarization of the initially excited excitons. This means that the exciton spin is robust enough to govern the alignment of Mn ions rather than the reverse. The experimentally observed formation of the spin-polarized EMPs requires then that the EMP formation time in CdMnTe QDs is comparable with or smaller than the spin relaxation time of the exciton. Moreover, the strong circular polarization of the CdMnTe QD emission seen in Fig. 1b means that, in contrast to DMS structures with higher dimensionality [5,6], the spin relaxation time of the EMP is significantly longer than the exciton recombination time in these QDs (300ps [15]).

The results obtained for different sample temperatures (see Fig, 2) show a gradual decrease of the polarization of the CdMnTe QD emission with increasing temperature. From an extrapolation of the temperature-dependent polarization of the first LO phonon replica we find that the emission becomes unpolarized around T~170 K. Previously, EMPs in both bulk DMSs [17] and DMS-based QWs [18] were observed to be stable only up to tens of degrees, while more strongly localized acceptor bound magnetic polarons persist up to 100 K [19]. Providing a localizing potential significantly improves the thermal stability of the magnetic polaron. EMPs localized in magnetic QDs can be viewed as a natural extension to these results. Theoretical calculations indicate that EMPs in magnetic QDs may exist up to nearly 100 K, depending on the dot size and Mn content [11]. Previous experiments performed for relatively large (25 nm in a diameter) CdMnSe QDs have shown stable zero-dimensional EMPs up to 30 K [8]. Stronger spatial confinement of EMPs in significantly smaller CdMnTe dots (~ 5 nm in a diameter) should enhance their thermal stability. Indeed, the strong polarization of emission observed at such high temperatures (see Fig. 2) indicates that the strong zero-dimensional confinement present in CdMnTe QDs significantly increases the thermal stability of EMPs.

In summary, we show that DMS self-assembled CdMnTe QDs can be polarized at *zero magnetic field* by optically aligning Mn ions within the dots and forming strongly localized magnetic polarons. The strong spatial confinement prevents the Mn ions from depolarizing, and as a result the emission from the CdMnTe self-assembled QDs shows predominant circular polarization for T>120 K.

**Acknowledgements**

The work was supported by the National Science Foundation through 9975655 and 0071797 (USA) the State Committee for Scientific Research under Grant PBZ-KBN-044/P03/2001 (Poland).

**Figure Captions**

**FIG 1.** Resonantly excited PL of (a) CdTe QDs and (b) CdMnTe QDs measured at B=0T and T=4.2 K. For $\sigma^+$ polarized excitation both $\sigma^-$ (squares) and $\sigma^+$ (circles) polarized PL is presented. The shaded areas represent the non-resonantly excited PL and the arrows mark the laser excitation energy.

**FIG 2.** Temperature dependence of the PL polarization at B=0 T for CdMnTe QDs under both $\sigma^-$ and $\sigma^+$ circularly polarized excitations. The solid line shows the data measured for CdTe QDs.

**FIG 3.** Schematics of the process of building up the spin polarization of magnetic polarons in CdMnTe QDs: (a) circularly polarized photon creates an electron-hole pair in a single QD, so that (b) the exciton is formed (shaded area). The Mn ions localized within the DMS QD align (c) according to the exciton spin polarization; as the result the emitted photon (d) maintains the polarization of the initial photon.

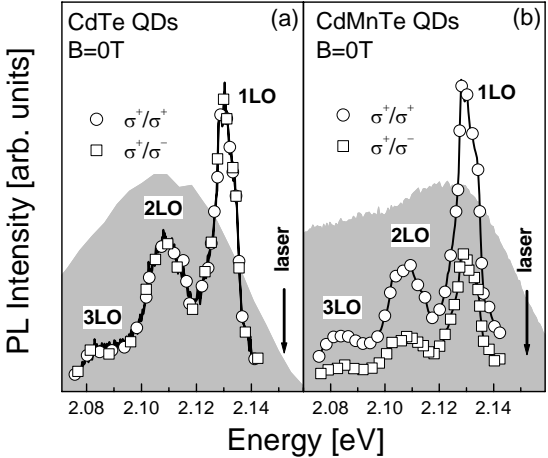

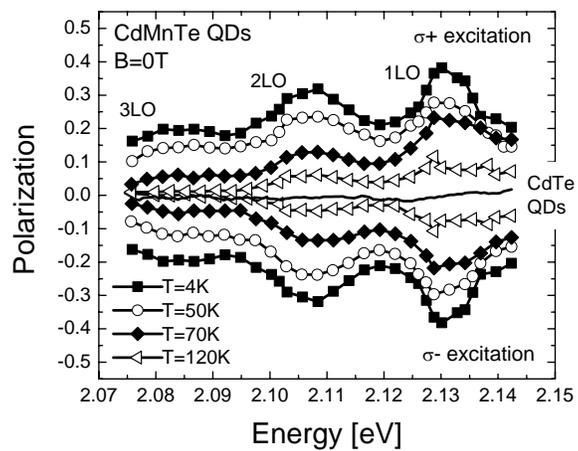

S. Mackowski *et al.*, Figure 3

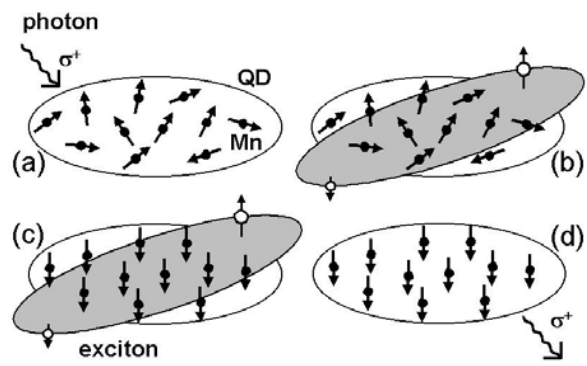

13